\documentclass[journal]{IEEEtran}

\usepackage{graphicx}
\usepackage{dcolumn}
\usepackage{bm}
\usepackage{booktabs}

\usepackage{xcolor}
\definecolor{OliveGreen}{RGB}{128,128,0}
\definecolor{Altrosa}{HTML}{E0707C}
\definecolor{rot}{RGB}{255,0,0}

\newcommand{\ttt}[1]{{\tt #1}}

\begin{document}

\newcommand{\citep}[1]{\cite{#1}}
\newcommand{\citet}[1]{\cite{#1}}


%
%
%
%
\title{Addressing Class Imbalance in Classification Problems of Noisy Signals by using Fourier Transform Surrogates}
%
%
\author{Justus T.~C.~Schwabedal, John C.~Snyder, Ayse Cakmak, Shamim Nemati, Gari D.~Clifford
\thanks{J.~Schwabedal, S.~Nemati, and Ayse Cakmak, and G.~Clifford are with the Department of Biomedical Informatics, Emory University, GA, USA}
\thanks{Manuscript recieved XXX, revised XXX}}

\maketitle

\begin{abstract}
Randomizing the Fourier-transform (FT) phases of temporal-spatial
data generates surrogates that approximate examples from the data-generating distribution.
We propose such FT surrogates as a novel tool to augment and analyze
training of neural networks and explore the approach in the example of sleep-stage classification.
By computing FT surrogates of raw EEG, EOG, and EMG signals of under-represented
sleep stages, we balanced the \ttt{CAPSLPDB} sleep database.
We then trained and tested a convolutional neural network for sleep
stage classification, and found that our surrogate-based augmentation
improved the mean F1-score by 7\%.
As another application of FT surrogates, we formulated an approach to compute
saliency maps for individual sleep epochs.  The visualization is based on the
response of inferred class probabilities under replacement of short data segments
by partial surrogates.
To quantify how well the distributions of the surrogates and the original data
match, we evaluated a trained classifier on surrogates of correctly
classified examples, and summarized these conditional predictions in a confusion
matrix.  We show how such conditional confusion matrices can qualitatively explain
the performance of surrogates in class balancing. 
The FT-surrogate augmentation approach may improve classification on noisy
signals if carefully adapted to the data distribution under analysis.
\end{abstract}

\begin{IEEEkeywords}
saliency maps, deep machine learning, class imbalance, sleep staging
\end{IEEEkeywords}

\section{Introduction}
\label{sec:intro}
Classification problems in biomedical signals are often imbalanced by one or more
orders of magnitude.  For example, epileptic seizures are rare minute-long events
that interrupt hours, days or even weeks of apparently normal cortical activity
in the electroencephalogram (EEG) \citep{mormann2006seizure-prediction}.
As another example, certain transitional sleep stages, such as S1 and S3, are
underrepresented with respect to more stable stages such as wakefulness or Rapid
Eye Movement (REM) sleep \citep{carskadon2005sleep-rev}.
Rare events, such as a possibly fatal \textit{status epilepticus} or sleep-onset
REM indicative of narcolepsy, are especially important in the biomedical realm.
It it is therefore imperative that such underrepresented classes are not swamped
by the more prevalent ones.
Classification algorithms such as logistic regression, support vector machines,
and random forest models can be extended to incorporate class imbalances in their cost
function structure (see Haixiang \textit{et al.} \citet{haixiang2017imbalance-rev} for
a review on class balancing). However by design, such extensions are aimed at general
applicability, offering only little flexibility to incorporate domain specific knowledge.
In mini-batch-based methods of deep learning, the imbalanced class distribution is typically
equilibrated by discarding examples of prevalent classes, or repeating those in the
minority.  A ten-fold up-sampling may, however, lead to partial over-fitting
whereas under-sampling unsatisfactorily discards vast amounts of valuable data.
Instead it has been proposed to sample from inferred distributions of minority classes 
\citep{chawla2002smote}.
In principle, deep generative models, in particular generative adversarial networks, can be used to
approximate examples from these distributions \citep{Mirza2014cgan,antoniou2017dagan}.  However,
these methods are very data-hungry, and we believe they will likely fail to generate
a variety of examples of rare classes in a dataset.
Class imbalance problem has also been addressed in the development of automatic sleep-staging systems
(see Aboaloayon \textit{et al.} \citet{aboalayon2016staging-rev} for a survey on such systems).
However in the few deep learning approaches we found, classes were balanced by either discarding data
\citep{tsinalis2016staging,dong2017staging}, or by up-sampling through repetitions of
data \citep{supratak2017staging}.  Both approaches introduce biases in the predictions,
either falsely pointing away from abnormality or falsely predicting illnesses.
Possible remedies can come from domain-knowledge based models of the rare class, which can either
be based in physical understanding of the biophysical process that generates the observations, or
through a statistical approach. 
In this article, we discuss up-sampling based on Fourier-Transform (FT)
surrogates \citep{schreiber2000surrogate-rev}.  We further describe a surrogate-based method
to construct saliency maps for a trained classifier.  Specifically, we
measure the response of inferred class probabilities under a surrogate replacement of short
data segments.
Our method adds to the several techniques that have been developed for image data, such as
methods of gradient ascent, saliency maps \citep{simonyan2013saliency}, and deconvolution
networks \citep{zeiler2014deconvolution}.
Previously the surrogate approach has been developed to test the hypothesis that a signal has
been generated by a linear stationary stochastic process, and has been
previously applied to EEG signals in this context \citet{andrzejak2001surrogates}.
However, we present a very different utilization in exploring the question of how
surrogates can facilitate machine learning, both, in training classifiers and in
analyzing what these learn to recognize.
%

%
%
%
%

%
\section{Methods}
\subsection{Surrogates based on the Fourier Transform}
\label{sec:fts}
The complex Fourier components $s_n$ of a signal $x_n$ can be decomposed into amplitudes $a_n$ and
phases $\varphi_n$ as $s_n=a_ne^{i\varphi_n}$.  Sample sequences of \textit{stationary linear
random processes} are uniquely defined by the Fourier amplitudes $a_n$, whereas their Fourier
phases are random numbers in the interval $[0,2\pi)$.  Under this assumption, we can draw a
new sequence $y_n$ that is statistically independent from $x_n$ while representing the same
generating distribution as first demonstrated by Theiler \citet{theiler1992surrogates}.  We simply
replace the Fourier phases of $s_n$ by new random numbers from the interval $[0,2\pi)$, and
apply the inverse Fourier transform.

\begin{figure*}[htb]
  \centering
  \includegraphics[width=0.79\linewidth]{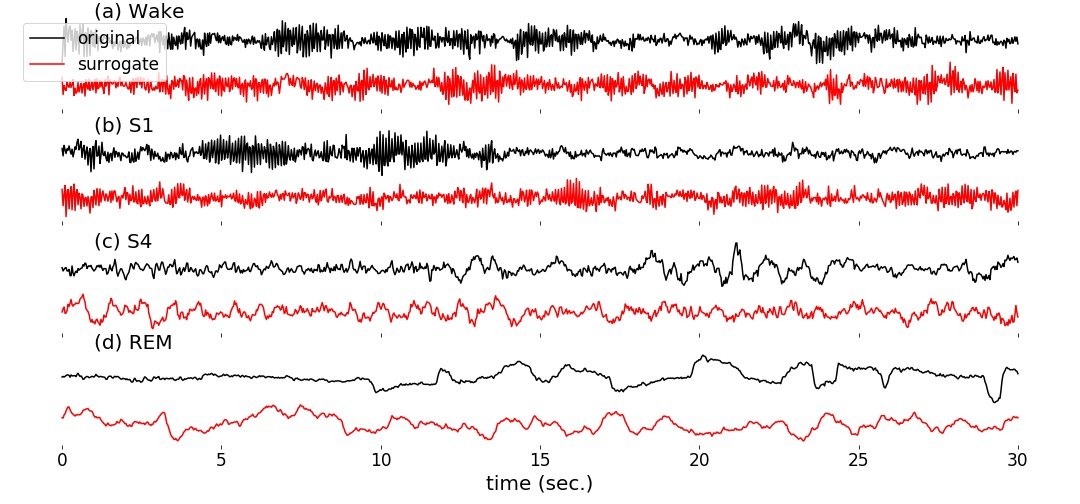}
  \caption{\textbf{Examples of FT surrogates.} We show four 30-second signals of EEG (a-c) and
  EOG (d) from \ttt{CAPSLPDB} recorded during different stages of sleep (indicated).  Each signal
  (black line) is counterposed with a representation of its FT surrogates (red line).  Panel~(b)
  illustrates the effect of non-stationarity on the technique, Panel~(d) that of nonlinearity.}
  \label{fig:surrogate-examples}
\end{figure*}

Under the assumptions of linearity and stationarity, we use this FT-surrogate method
to generate new independent samples of the sleep database analyzed here.  In Fig.~\ref{fig:surrogate-examples},
we show examples of EEG segments together with examples of their FT surrogates.
Example~(a) is dominated by EEG alpha waves centered around 10-Hz, wherein for example~(b), such alpha waves
are only visible in the segment's first half.  Comparing their surrogates allows us to understand better
the effect of nonstationarity on the FT surrogate technique.  While the surrogate represents the
data in example~(a) visually well, surrogate~(b) does not show a strong localization of the alpha
waves to a particular subsection.  The power in this band is smeared across the whole surrogate
segment thus leading to a very different visual appearance. 
Schreiber \textit{et al.} \citet{schreiber1996iaaft} extended the FT-surrogate method to simultaneously model the
time-domain amplitude distribution $P(x_n)$ in addition to the Fourier-amplitude
distribution $P(s_n)$ of the original signal.  In short, their algorithm starts by computing
a regular FT surrogate.  Next, the time-domain distribution of the surrogate is replaced
by the original one.  Then,the adjusted surrogate is Fourier transformed again and the Fourier
distribution is replaced by the original one.  The last two steps are repeated
iteratively until the time-domain distribution converges sufficiently.
Accordingly, these surrogates are called \textit{iterative amplitude-adjusted FT (IAAFT) surrogates}.
\subsection{Polysomnographic Database}
\label{sec:dataset}
\begin{figure}[h]
  \centering
  \includegraphics[width=0.79\linewidth]{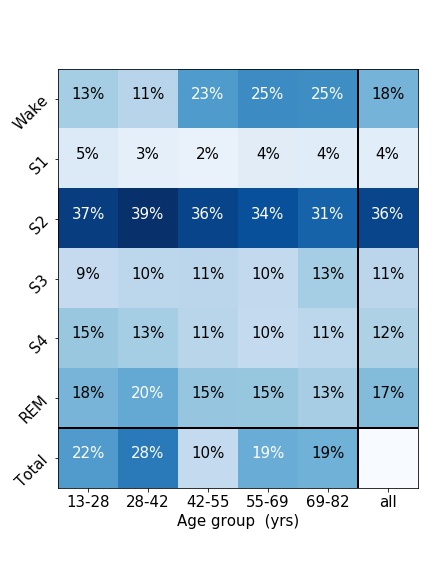}
  \vspace*{-0.8cm}
  \caption{\textbf{Stage distribution across age groups.}  The relative histogram displays
  the distribution of all 107,738 epochs stratified by age and sleep stage.}
  \label{fig:stage_distribution}
\end{figure}

We processed the \ttt{CAPSLPDB} sleep database consisting of 101 overnight polysomnographies (PSGs)
\citep{goldberger2000physiobank, terzano2001capslpdb}.  Each recording contained about eight hours
of multichannel recordings and sleep-stage annotations scored by an expert according to
R\&K~68 rules~\citep{RK68}.
We did not take into account recordings \ttt{rbd11}, \ttt{brux1}\footnote{The score of \ttt{brux1}
was recovered after the analysis.}, and \ttt{nfle27} because of missing sleep scores, and
\ttt{n4}, \ttt{n8}, \ttt{n12}, and \ttt{n16} because these only contained EEG channels.
The remainder of recordings were divided into five equidistant age bins.  The division was based
on the data distribution.
Each record had been divided in 30-second intervals each assigned one of the sleep stages
\ttt{Wake}, \ttt{S1}, \ttt{S2}, \ttt{S3}, \ttt{S4}, \ttt{REM}, or \ttt{MT} by an expert
sleep technician.  In Fig.~\ref{fig:stage_distribution}, we summarize the distribution of
stages stratified by age groups.  We ignored stage~\ttt{MT} which occurred only $685$~times. 
In each age group, stage~\ttt{S1} was least well represented, averaging at $4\%$ across
all groups.  The fraction of stage~\ttt{Wake} increased with age, and meanwhile the
fraction of \ttt{S4} and \ttt{REM} decreased.
From all available channels in each recording, we select a subset including two EEG channels,
one EOG, and one EMG channel, which is a maximal subset included in all recordings.  The available
EEG channels were also heterogeneous regarding the recording site and derivation.  They were
selected with a preference list (numbers included EEG1 and EEG2):
\ttt{F3-C3} ($n=67$),
\ttt{P3-O1} ($67$),
\ttt{C4-M1} ($29$),
\ttt{F4-C4} ($26$),
\ttt{C3-M2} ($6$),
\ttt{O2-M1} ($3$),
\ttt{P3-Cz} ($2$),
\ttt{F7-Cz} ($2$).
We resampled all signals to $32$~Hz after applying a $13$~Hz $4$-th order
Butterworth low-pass filter to reduce aliasing.

Alongside different stages of sleep, aging is also known to correlate
with characteristic EEG patterns.  The co-variation leads to an implicit class
under-representation of wakefulness in young, for example.  Moreover, the
database consists of individuals suffering from various diseases or disorders
that are represented differentially across age, and available channels reflect
to some extend the disease-specific investigation:  records of young
nocturnal-frontal-lobe-epilepsy patients included more EEG channels than
regular PSGs, for example.
We do not attempt to address all of these sources of class imbalance within
the scope of this article, because on this level of detail, the present
database is too small.
%

%
%

\subsection{Network Architecture and Training}

We explored a convolutional neural-network architecture as a deep learning
model for our sleep database.  The goal was to optimize the F1-score
for all six classes across the different age groups.  We used Google Cloud's ml-engine
infrastructure for all computations including Bayesian hyper-parameter optimization.

Our architecture takes as input 30-second raw sequences of two EEG, one EMG, and one EOG
channel as one example, and outputs soft maximum-based probabilities for the six classes
\ttt{Wake}, \ttt{S1}, \ttt{S2}, \ttt{S3}, \ttt{S4}, and \ttt{REM}.  Two parts constitute
our network architecture:  first each channel is processed by dedicated neural networks
only operating on that one channel; and second, their outputs are merged to process
interrelation among channels.  Note, that the 4-channel input would suffice for
sleep-scoring experts to deduce sleep stages.  The network architecture is
summarized in detail in Tab.~\ref{tab:architecture}.

\noindent\textbf{Channel pipes.}  In the first stage each channel is processed with a pipe of
one-dimensional convolutional layers.  While all pipes share the same architecture,
each channel type has its own parameters, i.e.~the two EEG channels share the same parameters.
We choose parameter sharing across EEG channels because the heterogeneity in our dataset
prohibited to train dedicated channels for specific electrode locations.  Choosing
the same pipe architecture for each channel facilitated joining their outputs in the second
stage.  After each convolutional layer we apply dropout with $p=0.33$.  The Scale layer was
initialized with a factor $0.05~\mu$V$^{-1}$.  Biases were initialized as zero, and
weights were initialized drawing from standard Glorot-uniform distributions \citep{glorot2010glorot}.

\noindent\textbf{Joined pipe.}  In the second stage, outputs of the first stage are stacked to form
a $(n, 4, f)$-dimensional tensor, where $f$ is the number of filters, and $n$ the length of 
each of the four joined sequences.  A two-dimensional convolution layer is applied to the result,
followed by two dense layers and the six-neuron soft-max layer to be matched with class probabilities.
After the first two dense layers, we apply dropout with $p=0.015$.  Biases were initialized as zero, and
weights were initialized drawing from a Glorot uniform distribution.
\begin{table}
  \caption{\textbf{Neural Network architecture.} F: number of filters,
  W: dimension of each filter, S: Stride parameter, ReLU: Rectified linear unit.}
  \label{tab:architecture}
  \begin{center}
  \begin{small}
  \begin{sc}
    \begin{tabular}{lll}
     \toprule
     \textbf{Name} 	     & \textbf{Description} & \textbf{Output} \\
     \midrule
     \multicolumn{3}{l}{Channel-pipe architecture} \\
     \multicolumn{3}{l}{each with 32,936 trainable parameters.} \\
     \midrule
     Input    & 30-second signal & $960$ \\
     Scale    & Scalar rescaling & $960$ \\
     Conv1D   & W: 16, F: 16, ReLU& $960\times16$ \\
     MaxPool  & W: 3, S: 2 & $480\times16$ \\
     Conv1D   & W: 19, F: 19, ReLU & $480\times19$ \\
     MaxPool  & W: 3, S: 2 & $240\times19$ \\
     Conv1D   & W: 23, F: 23, ReLU & $240\times23$ \\
     MaxPool  & W: 3, S: 2 & $120\times23$ \\
     Conv1D   & W: 27 , F: 27, ReLU & $120\times27$ \\
     MaxPool  & W: 3, S: 2 & $60\times27$ \\
     \midrule
     \multicolumn{3}{l}{Joined-pipe architecture} \\
     \multicolumn{3}{l}{with 64,371 trainable parameters.} \\
     \midrule
     Input    & Output of channel pipes & $60\times4\times27$ \\
     Conv2D   & W: $20\times4$, F: 10, ReLU & $41\times1\times10$ \\
     Dense    & 85 neurons, ReLU & $85$ \\
     Dense    & 85 neurons, ReLU & $85$ \\
     Dense    & 6 neurons, soft-max & $6$ \\
     \bottomrule
    \end{tabular}
  \end{sc}
  \end{small}
  \end{center}
\end{table}
We trained the network on $7000$ mini-batches of $128$ examples, and using an RMS-Prop
optimization algorithm with a learning rate of $0.0016$, a decay parameter of $0.9$, and
no momentum \citep{tieleman2012rmsprop}.  The number of steps was chosen through our
experience of visually inspecting validation and training loss, and assured that these
quantities always reached stable values.  

\subsection{Validation Split and Data Sampling}

We split off a validation set from the database by holding one recording back from
each age group.  On these five recordings, we validated an instance of a neural network
which we trained on the training set consisting of all other records. 
In a \textit{5-fold cross validation}, we split the database (and trained networks)
five times, each with different validation recordings.  This yielded a total validation
set of five recordings from each age group, i.e.~a total of 25 recordings.
During training, we sorted the training set by stage label for up-sampling and augmentation
which we controlled by two parameters $\beta\in[0,1]$ and $\alpha\in[0,1]$.
As a last step, we shuffled the processed training set to randomly group examples into
mini-batches of size $128$.
\noindent\textbf{Up-sampling.}
We computed the number of repetitions $n_c$ of under-represented class $c$ necessary to match the number
of the most frequent class.  We then multiplied $n_c$ by a factor~$\beta$, and added
a corresponding number of random repetitions to the training set.  The factor allowed us to control up-sampling.
For the presented results, we set $\beta=0.9$.

\noindent\textbf{Augmentation.}
Each channel in the repeated examples were replaced by FT surrogates with probability
$\alpha$.  That means for a given repeated example that only some of its channels could
be augmented by surrogate replacements.
With this publication we provide the preprocessed database and scripts reproducing our
results.\footnote{\ttt{https://github.com/cliffordlab/sleep-convolutions-tf}}

\section{Results}
\subsection{Training with FT Surrogate-based Class Balancing}
\label{sec:surrogatesampling}
\begin{figure}[h]
  \includegraphics[width=0.85\linewidth]{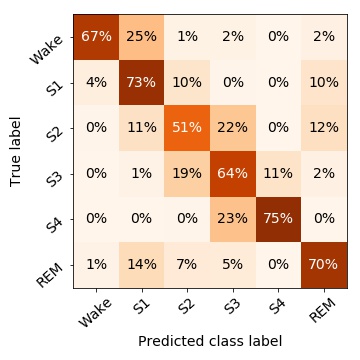}
  \caption{\textbf{Confusion matrix of test-set predictions.}
  We evaluated the network on examples from the test set and computed the
  fraction of labels of a certain class with respect to their predictions
  shown in this color-coded confusion-matrix representation.}
  \label{fig:confusion}
\end{figure}

\begin{figure}[h]
  \centering
  \includegraphics[width=0.99\linewidth]{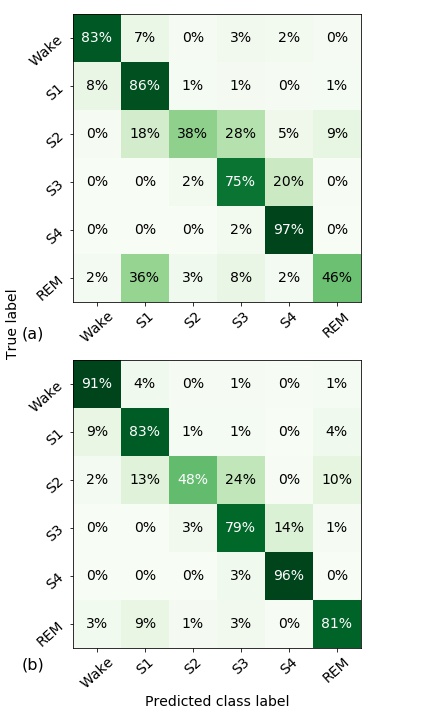}
  \caption{\textbf{Conditional confusion matrix.}  The correct predictions
  (diagonal in Fig.~\ref{fig:confusion}) were transformed to (a) FT surrogates,
  and (b) IAAFT surrogates, and than re-scored by the sleep-staging algorithm.
  The result is presented in respective conditional confusion matrices.}
  \label{fig:ccm}
\end{figure}

We started our analysis by training the feed-forward neural network model without replacing
any repeated signals by FT~surrogates ($\alpha=0$).  Our training did not show
considerable over-fitting as indicated by a close proximity of training- and test-set
accuracies.  Though not groundbreaking, our classification results on the five-fold
validation set shown in Fig.~\ref{fig:confusion} were within the range of previously
reported results for sleep-stage classification, especially for the very complex
\ttt{CAPSLPDB}.
We leveraged the trained model to investigate how well signals of different sleep stages
are represented by their FT~surrogates.   For each correctly predicted example, we
computed a surrogate and re-applied the classifier.  We analyzed the confusion matrix
for these surrogate labels conditioned on a correct original prediction.
As shown in Fig.~\ref{fig:ccm}(a), FT surrogates of stages \ttt{Wake}, \ttt{S1},
and \ttt{S4} were predicted to be from the correct class with probabilities larger than 80\%,
whereas surrogates of \ttt{S2} and \ttt{REM} showed lowest conditional accuracies.
Comparing the off-diagonal matrix elements, we found that \ttt{S1}-surrogates
are more often miss-classified as \ttt{Wake}, \ttt{S2} as \ttt{S3}, \ttt{S3}
as \ttt{S4}, and \ttt{REM} as \ttt{S1}.
Exemplary, the miss-classification \ttt{S1}$\to$\ttt{Wake} may be explained
by the redistribution of non-stationary bursts of alpha oscillations when
drawing a surrogate as visible in Fig.~\ref{fig:surrogate-examples}(b):
in the surrogate, the alpha rhythm appears in more than 50\% of the segment
thus making a classification \ttt{Wake} more likely by eye and by algorithm.
We hypothesize that miss-classifications \ttt{S2}$\to$\ttt{S3}, \ttt{S3}$\to$\ttt{S4},
and \ttt{REM}$\to$\ttt{S1} are also due to non-stationarities, i.e., K-complexes,
and bursts of delta waves or rapid eye movements.
We also evaluated the conditional confusion matrix when replacing the original
correctly predicted examples with IAAFT surrogates, as shown in Fig.~\ref{fig:ccm}(b).
Comparing the conditional accuracies of FT and IAAFT surrogates, we observed
that the latter were equal or better predicted for all stages except \ttt{S1}.
Standard deviations in conditional confusion values were around 1\%.
%

%
%

\begin{figure}[h]
  \includegraphics[width=0.99\linewidth]{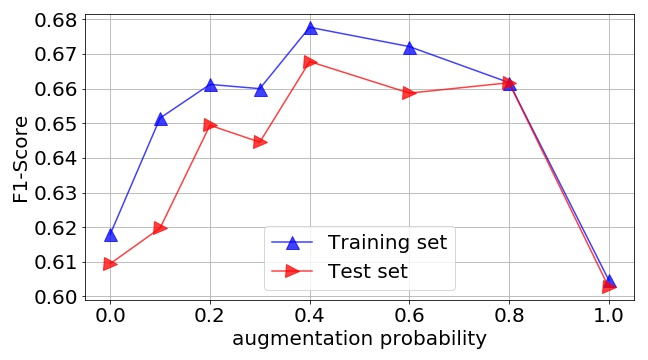}
  \caption{\textbf{Average F1-score versus augmentation probability.}  The average F1-score depending
  on surrogate augmentation probability $\alpha$ shows a distinct maximum, both for the test set
  as well as for the training set.}
  \label{fig:augment}
\end{figure}

\begin{figure}[h]
  \includegraphics[width=0.99\linewidth]{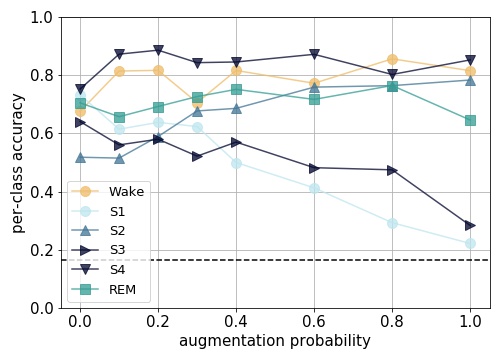}
  \caption{\textbf{Per-class accuracy versus augmentation probability.}  The per-class accuracy depending
  on surrogate augmentation probability $\alpha$ shows slightly increasing patterns for stages \ttt{Wake},
  \ttt{S2}, \ttt{S4}, and \ttt{REM}, and sharply decreasing patterns for stages \ttt{S1}, and \ttt{S3}.  This
  discrepancy explains to a certain extend the convex F1-score dependence (cf.~Fig.~\ref{fig:augment}).}
  \label{fig:class-acc}
\end{figure}

%
Next, we increased the augmentation probability $\alpha$ to values between $0$ and $1$,
thus replacing fractions of up-sampled signals by FT surrogates.  At each $\alpha$,
we performed our scheme of five-fold cross-validation and observed how prediction probabilities
changed.
We found a consistent maximum of the F1-score at about $\alpha=0.4$
(cf.~Fig.~\ref{fig:augment}).
The convex dependence of the F1-score on $\alpha$ can be better understood
when decomposing the measure into its constituent per-class accuracies summarized
in Fig.~\ref{fig:class-acc}.  While the accuracy of stages \ttt{Wake}, \ttt{S2},
and \ttt{S4} slightly increase, the \ttt{S1}- and \ttt{S3}-accuracies rapidly decrease
towards zero beyond $\alpha>0.4$.  These two opposing objectives create
the quantitative compromise exhibited as a non-trivial maximum in the
$\alpha$-dependence of the F1-score.
Notice that the accuracy of stage \ttt{S2} showed the greatest benefit of surrogate-based up-sampling,
of which no surrogates were created.
Unfortunately, we were not yet able to evaluate and compare IAAFT surrogates with these results
due to temporal and budget constraints.

\subsection{Partial FT Surrogates to Analyze Class Probabilities}
\label{sec:pfts}

\begin{figure}[htb]
  \centering
  \includegraphics[width=0.99\linewidth]{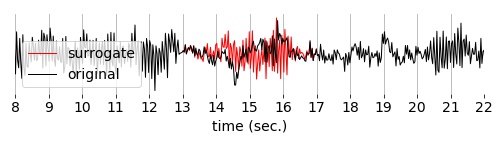}
  \caption{\textbf{Example of a partial FT surrogate.}  A 4-second subsection of an EEG signal
  recorded during stage \ttt{Wake} is shown (black line), together with a partial FT surrogate
  (red line).  The partial surrogate replaced the anomaly in the segment between second 13 and 17. 
  Note that the surrogate dominantly contains the $\sim$ 10-Hz alpha waves also visible in the rest of
  the signal.}
  \label{fig:pfts}
\end{figure}

Based on FT surrogates we propose a novel technique to create saliency maps from which we
can read out the relative importance of a subsection of a signal for the predicted
class probabilities.
First, we selected a window length and a subset of channels in which we presumed to find
a relevant feature.  To query the relevance of the data at a given location in the epoch,
one could, naively, zero-out the subsection in question and observe how inferred
probabilities change.  However, imputing such quiescent periods can introduce class
biases; for example a very low-voltage \ttt{EMG} signal strongly indicates
\ttt{REM} sleep over other sleep stages.
Instead, we spliced out the signal window, and replaced the subsection with an
FT surrogate generated from the remainder of the signal under analysis as
visualized in Fig.~\ref{fig:pfts}.  All splicing was performed smoothly
by cosine half-wave interpolation of $0.5$-second overlaps.
\begin{figure*}[hbt]
  \centering
  \includegraphics[width=0.78\linewidth]{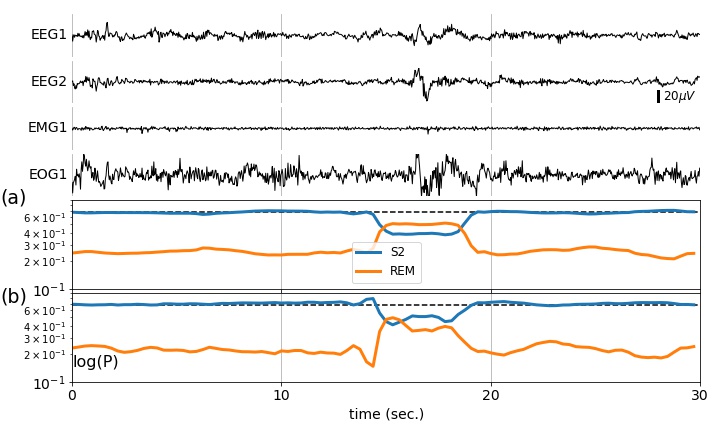}
  \caption{\textbf{Partial FT surrogate analysis.}  This 30-second \ttt{REM} epoch was miss-classified
  as \ttt{S2} with $P(\tt{S2})=68\%$ (black dashed line).  The suspicion was that the K-complex at about
  17~seconds caused the miss-classification.  The used classifier was trained with $\alpha=0.4$.
  \textbf{(a)} We analyzed the epoch with the partial
  FT-surrogate technique to both \ttt{EEG} signals using a 5-second-long moving window with an
  overlap of $0.5$~seconds, and $500$ surrogate replacements.  The averaged probability of \ttt{S2}
  and \ttt{REM} change as a function of the window location.
  The temporary reversal of probabilities indicates that the K-complex at about 17 seconds caused the
  miss-classification. \textbf{(b)} The surrogate approach is counterposed with simple zeroing out,
  in which an equivalent 5-second window
  is (smoothly) replaced by zeros.  This naive approach also shows a reversal of probabilities, but
  at the wrong position.  Note that there was no offset in the signal.}
  \label{fig:example}
\end{figure*} 
For a given window location, the partial surrogate replacement was performed multiple times. 
For each replacement, the epoch was then processed by our sleep-staging algorithm and
the class probabilities were recorded.  Finally, we averaged these class probabilities over
the independent replacements.  The averaged probabilities as a function of
the window position yielded a saliency map that described the relevance of localized features
for the classification result found in for a specific example.
We demonstrate the partial FT surrogate technique with an example epoch of stage~REM that has been misclassified
as stage~S2 by our algorithm (cf.~Fig.~\ref{fig:example}).  In the latter half of the example,
there is a K-complex visible in both EEG-channels and the EOG-channel, which according to the rules leads
to a stage change to S2 in the following epoch.  Had it occurred in the earlier half of the example, the
example would have been scored as S2.
We analyzed this epoch using our partial-surrogate method (cf.~Fig.~\ref{fig:example}(a)), and counterposed
the result with naive zeroing out of equivalent subsections (cf.~Fig.~\ref{fig:example}(b)).
The prediction probabilities of sleep states S2 and REM crossed or reversed as the surrogate replacement 5-second window slides across the location of
the K-complex.  The probabilities also reversed for the zero-out method, however, not concurrently with 
the visually identified event.

\section{Discussion}
\label{sec:discussion}
We explored two applications of Fourier transform (FT) surrogates to sleep stage classification:
we analyzed how up-sampling minority examples with FT surrogates affects the prediction scores.
Furthermore, we described a method of saliency maps based on partial FT surrogates that allow us to
analyze how individual class probabilities depend on subsections of the signals.%
The convex dependence of the F1-score on the augmentation probability indicates a
possible benefit of surrogate-based up-sampling.  However, this might not be the case for all
class labels equally.  Increases in the \ttt{S2}-accuracy seemed to be at the expense of
stages \ttt{S1} and \ttt{S3} for larger values of $\alpha$.  Based on these results,
we hypothesize that the effect of surrogate augmentation on an individual class accuracy
does not directly depend on their conditional prediction accuracies, which are on the diagonal
of the conditional confusion matrix (cf.~Fig.~\ref{fig:ccm}(a));  instead,
augmentation may introduce mixing between class labels indicated by a large off-diagonal
element upon which the accuracy of one of the mixed labels will dominate.  Accordingly,
we hypothesize the accuracy increase of \ttt{S2} and \ttt{REM} to be at the expense of
classification accuracy of \ttt{S1}, and the increase in accuracy for \ttt{S2} and
\ttt{S4} at the expense of classification accuracy of \ttt{S3}.
The conditional confusion matrix of IAAFT surrogates exhibit higher accuracy and lower
off-diagonal elements indicating mixing of labels (cf.~Fig.~\ref{fig:ccm}(b)).
One interpretation of the results is that IAAFT surrogates are able to model the data distribution
more accurately; on the other hand,
the results are also consistent with the data distribution to be highly collapsed into
regions that are well predicted by our algorithm.  While the former would suggest benefits
of using IAAFT over simple surrogates, the latter would mean that using IAAFT would increase
the tendency to over-fit the data.  To date, we understand little about the topological
properties of the IAAFT distribution and therefore it is hard to reason which effect
will dominate.
Therefore, it would be interesting to see how training with IAAFT surrogates impacts
accuracy scores in this and other examples of biomedical data analysis.
Specifically we predict from our hypothesis, that augmentation with IAAFT surrogates
will have a less negative impact on the \ttt{S1} classification accuracy.
Partial surrogate analysis is not restricted to neural-network based or other differentiable
classifiers as these saliency maps are created purely by controlling input and output
probabilities.
Also, the technique, aimed at transient signal features, does not greatly suffer from
the requirement of stationarity since the replaced subsections are of lengths at which
EEG signals are approximately statistically stationary.
However, features without temporal localization cannot be delineated with our
technique.  For example, a constant alpha-wave background will not be detected
to distinguish \ttt{Wake} from \ttt{S1} because the surrogate replacement will
also contain alpha waves (compare~Fig.~\ref{fig:surrogate-examples}(a) and (b)).
Such features are more likely to be highlighted by gradient-based saliency maps, and when
training on a wavelet representation of the signal as data input.
The example shown in Fig.~\ref{fig:example} highlights the strength of our technique,
where it allowed us to gather evidence that our sleep-staging algorithm learned
about the existence of K-complexes and their relevance of distinguishing between
\ttt{REM} and~\ttt{S2} (cf.~Fig.~\ref{fig:example}).  This was particularly unclear
given the relatively poor accuracy of the classifier.
%

%
%
We conclude from the present work that the ability to draw independent examples
from the data distribution is important in training, analysis, and validation of
deep machine-learning models.  As in this work, such examples can be used to balance
and augment a database to achieve better generalization, and to understand which
statistical properties of data are instrumental to black-box learning algorithms
to make predictions.
Unless the database is large enough to train a deep generative model that mimics the data
distribution, it is necessary to build the generator from a strong set of constraints rooted
in specific domain knowledge.  This is especially the case for under-represented classes
for which we do not have a lot of data.  Usage of FT surrogates is constrained to stationary
linear random data as the current work illustrates.  For IAAFT surrogates we cannot formulate the
precise constraints.  In the future, it may also be helpful to query mechanism-based models
to generate surrogates in situations, particularly for nonlinear signals, that are not well
represented by FT-based surrogates, such as electrocardiograms.
In the future, we plan to adopt our approach to identify ambiguous or mislabeled data
which are often mislabeled for two general reasons:
natural inter- and intra-observer variability for transitional epochs, and errors due to
quantization or coarse windowing of data.  Although the issue of only moderate inter- and
intra-rater agreement levels is a known issue in sleep stage labeling \cite{DANKER-HOPFE2009},
the latter issue is a particularly under-explored problem in sleep stage classification.
In particular we plan to use partial FT surrogate analysis to identify epochs ambiguous due
to short transient events.  The ability to programatically exclude such edge cases from
a training may enhance the efficacy of sleep-stage classification.
\section*{Acknowledgements}
This research is supported in part by funding from the James S. McDonnell
Foundation, Grant 220020484 (http://www.jsmf.org), the Rett Syndrome Research
Trust and Emory University and the National Science Foundation Grant 1636933
(BD Spokes: SPOKE: SOUTH: Large-Scale Medical Informatics for Patient Care
Coordination and Engagement). Dr. Nemati's research is funded through an NIH
Early Career Development Award in Biomedical Big Data science
(1K01ES025445-01A1).  This work was partially funded by NSF grant 1822378.
%



\end{document}